\pdfoutput=1  
\def\nk{n_\mathrm{b}}
\def\acap{\\ \nonumber \\}

\def\Pb{P_\mathrm{b}}
\def\rfr#1{Equation\,(\ref{#1})}
\def\rfrs#1#2{Equations\,(\ref{#1})--(\ref{#2})}
\def\Rfr#1{Equation\,(\ref{#1})}

\def\derp#1#2{\rp{\partial{#1}}{\partial{#2}}}
\def\dert#1#2{\frac{{{\textrm{d}}}{#1}}{{{\textrm{d}}}{#2}}}

\def\eqi{\begin{equation}}
\def\eqf{\end{equation}}
\def\rp#1#2{\frac{#1}{#2}}
\def\lb#1{\label{#1}}

\def\ton#1{\left(#1\right)}
\def\qua#1{\left[#1\right]}
\def\grf#1{\left\{#1\right\}}

\RequirePackage[2020-02-02]{latexrelease}
\documentclass{aastex631}
\usepackage{morefloats}
\usepackage[title]{appendix}
\usepackage{textcomp}
\usepackage{booktabs}
\usepackage{multirow}
\usepackage{rotating,tabularx}
\usepackage{float}
\usepackage{enumerate}
\usepackage{rotating}
\usepackage[polutonikogreek,english]{babel}
\usepackage{amsmath,starfont,textgreek,w-greek}
\usepackage[flushleft]{threeparttable}
\usepackage{amsthm}
\usepackage{bigints}
\usepackage{amscd}
\usepackage[mathlines]{lineno}
\usepackage{amssymb,dsfont}
\usepackage{graphicx,epsfig}
\usepackage[nointegrals]{wasysym}
\usepackage[caption=false]{subfig}

\allowdisplaybreaks

\makeatletter
 \DeclareRobustCommand\ref{%
    \@ifstar\@refstar\T@ref
  }%
  \DeclareRobustCommand\pageref{%
    \@ifstar\@pagerefstar\T@pageref
  }%
 \makeatother

\begin{document}

\title{On the Euler--type gravitomagnetic orbital effects in the field of a precessing body}

\shortauthors{L. Iorio}

\author[0000-0003-4949-2694]{Lorenzo Iorio}
\affiliation{Ministero dell' Istruzione e del Merito. Viale Unit\`{a} di Italia 68, I-70125, Bari (BA),
Italy}

\email{lorenzo.iorio@libero.it}

\begin{abstract}
\textcolor{black}{To the first post--Newtonian order, the gravitational action of mass--energy currents is encoded by the off--diagonal gravitomagnetic components of the spacetime metric tensor. If they are time--dependent, a further acceleration enters the equations of motion of a moving test particle. Let the source of the gravitational field be an isolated, massive body rigidly rotating whose spin angular momentum experiences a slow precessional motion. The impact of the aforementioned acceleration on the orbital motion of a test particle is analytically worked out in full generality.}
The resulting averaged rates of change are valid for any orbital configuration of the satellite; furthermore, they hold for an arbitrary orientation of the precessional velocity vector of the spin of the central object. In general, all the orbital elements, with the exception of the mean anomaly at epoch, undergo nonvanishing long--term variations which, in the case of the Juno spacecraft currently orbiting Jupiter and the double pulsar PSR J0737--3039 A/B turn out to be quite small. Such effects might become much more relevant in a star--supermassive black hole scenario; as an example, the relative change of the semimajor axis of a putative test particle  orbiting  a Kerr black hole as massive as the one at the Galactic Centre at, say, 100 Schwarzschild radii may amount up to about $7\%$ per year if the hole's spin precessional frequency is $10\%$ of the particle's orbital one.
\end{abstract}

\section{Introduction}
In an accelerated reference frame rigidly rotating with time--dependent angular velocity ${\boldsymbol{\Omega}}_\mathrm{L}\ton{t}$,
a particle located at position $\boldsymbol{r}$ experiences, among other things, also the fictitious Euler acceleration \citep{Eul1,Eul2,Eul3,Eul4}
\eqi
\textcolor{black}{{\boldsymbol{\mathcal{A}}}_\mathrm{E} = -\dert{{\boldsymbol{\Omega}}_\mathrm{L}} t\boldsymbol{\times}\boldsymbol{r},}\lb{AccEul}
\eqf
\textcolor{black}{which is often neglected}.
\textcolor{black}{On the basis of the equivalence principle, at the foundation of the General Theory of Relativity (GTR), one may expect that, within certain limits, an analogous acceleration of gravitational origin should act on a test particle orbiting a rotating body as seen in a local\footnote{\textcolor{black}{In the sense that its extension is assumed to be small enough to neglect residual tidal effects due to any external fields.}} inertial frame attached to the latter. As it will be shown explicitly in the following, it is just the case. After all, the Lense--Thirring (LT) acceleration \citep{1918PhyZ...19..156L,1984GReGr..16..711M} is the general relativistic counterpart, to the first post--Newtonian order (1pN), of the largely known fictitious Coriolis acceleration affecting the motion of an object referred to a rigidly rotating frame.
}

\textcolor{black}{The spin angular momentum $\boldsymbol J$ of an extended, rigidly rotating  body of mass $M$ is often displaced, to the Newtonian level, by the differential gravitational tugs exerted on different parts of its centrifugal bulge by other  distant masses. It is just the case of the precession and nutation of the Earth mainly due to the torques exerted by the Moon and the Sun \citep{PrecNut}.} Precessional motions of $\boldsymbol J$ occur also to the 1pN order when the body of interest moves in the deformed spacetime of other objects \citep{DamRu74,1975PhRvD..12..329B}.

Does the temporal variation of $\boldsymbol{J}$ directly affect the orbital motion of a test particle revolving about its spinning primary? \textcolor{black}{While the answer is negative to the Newtonian level, it is generally affirmative to the 1pN order.} If, on the one hand,  such general relativistic effects are usually quite small for ordinary restricted two--body systems, on the other hand, they might become relevant when supermassive black holes are involved; be that as it may,  it is worthwhile calculating them in full generality.

According to \citet[Eq. (2.2.49), pag. 56]{1991ercm.book.....B}, the 1pN equations of motion of a test particle contain, among other things, a time--varying gravitomagnetic acceleration which reads
\eqi
\boldsymbol{A} = c^2\derp{\boldsymbol{\mathfrak{h}}}{x^0} = c\derp{\boldsymbol{\mathfrak{h}}}{t}\textcolor{black}{.}\lb{Ace}
\eqf
\textcolor{black}{In \rfr{Ace},} $c$ is the speed of light in vacuum, $x^0:= ct$ is the time--like coordinate, and
\eqi
\boldsymbol{\mathfrak{h}}:=\grf{h_{01},h_{02},h_{03}}\lb{hac}
\eqf
is made of the  pN corrections $h_{0i},\,i=1,2,3$ to the otherwise vanishing off--diagonal components $\eta_{0i}=0,\,i=1,2,3$ of the Minkowskian spacetime metric tensor $\eta_{\mu\nu},\,\mu,\nu=0,1,2,3$. They are due to the mass--energy currents of the source of the gravitational field, and are conventionally dubbed as gravitomagnetic.
In general, their  effects  can formally  be described in terms of a gravitomagnetic field ${\boldsymbol{\mathrm{B}}}_\mathrm{g}$ which can be obtained from a gravitomagnetic potential vector \citep{2001rfg..conf..121M,Mash07}
\eqi
{\boldsymbol{\mathrm{A}}}_\mathrm{g} :=  \rp{c^2\boldsymbol{\mathfrak{h}}}{2} \lb{hacca}
\eqf
as
\eqi
\boldsymbol\nabla\boldsymbol\times{\boldsymbol{\mathrm{A}}}_\mathrm{g} = -\rp{{\boldsymbol{\mathrm{B}}}_\mathrm{g}}{2}.\lb{nablaA}
\eqf

In the case of an isolated, rigidly rotating body,
\rfr{hac} turn out to be \citep{2001rsgc.book.....R,2002NCimB.117..743R,2014grav.book.....P}
\eqi
\boldsymbol{\mathfrak{h}} = \rp{2G \boldsymbol J\boldsymbol\times\boldsymbol r}{c^3 r^3},\lb{gis}
\eqf
where $G$ is the Newtonian constant of gravitation, and $\boldsymbol{r}$ is the position vector of a test particle revolving about the central object, being $r$ their mutual distance. Thus, for such a source, the gravitomagnetic potential can be cast into the form
\eqi
{\boldsymbol{\mathrm{A}}}_\mathrm{g} = \rp{G\boldsymbol J\boldsymbol\times\boldsymbol r}{c r^3},\lb{Ag}
\eqf
and the resulting gravitomagnetic field is
\eqi
{\boldsymbol{\mathrm{B}}}_\mathrm{g} =  \rp{2G\qua{\boldsymbol{J} - \ton{\boldsymbol J\boldsymbol\cdot\boldsymbol{\hat{r}}}\boldsymbol{\hat{r}}}}{c r^3}.\lb{Bg}
\eqf

If $\boldsymbol{J}$ varies over time, \rfr{Ace} and \rfr{gis} yield
\eqi
\boldsymbol{A} = \rp{2G}{c^2 r^3}\derp{\boldsymbol{J}}{t}\boldsymbol\times\boldsymbol{r}.\lb{acc}
\eqf
In a body--fixed rotating frame with angular velocity ${\boldsymbol{\omega}}_0$, the dynamical Euler equations
\eqi
\dert{\boldsymbol{J}}t = \derp{\boldsymbol{J}}t + {\boldsymbol{\omega}}_0\boldsymbol\times\boldsymbol{J}\lb{eul}
\eqf
hold; since an inertial frame is assumed, \rfr{eul} reduces to
\eqi
\dert{\boldsymbol{J}}t = \derp{\boldsymbol{J}}t.\lb{part}
\eqf
According to \rfr{part}, \rfr{acc} becomes
\eqi
\boldsymbol{A} = \rp{2G}{c^2 r^2}\dert{\boldsymbol{J}}{t}\boldsymbol\times\boldsymbol{\hat{r}}.\lb{Accel}
\eqf

\textcolor{black}{Can the existence of \rfr{Accel} be guessed on the basis of the equivalence principle, in analogy with the fictitious Euler acceleration of \rfr{AccEul}?}
The gravitational analogue of the Larmor theorem \citep{1993PhLA..173..347M} tells that the geodesic motion of a test particle in a spatially uniform gravitomagnetic field ${\boldsymbol{\mathrm{B}}}_\mathrm{g}$, characterized by\footnote{It can be straightforwardly shown that \rfr{Olar2} fulfils \rfr{nablaA} by means of
$\boldsymbol\nabla\boldsymbol\times\ton{\boldsymbol{\mathrm{P}}\boldsymbol\times\boldsymbol{\mathrm{Q}}} = \boldsymbol{\mathrm{P}}\ton{\boldsymbol\nabla\boldsymbol\cdot\boldsymbol{\mathrm{Q}}} - \boldsymbol{\mathrm{Q}}\ton{\boldsymbol\nabla\boldsymbol\cdot\boldsymbol{\mathrm{P}}} + \ton{\boldsymbol{\mathrm{Q}}\boldsymbol\cdot\boldsymbol\nabla}\boldsymbol{\mathrm{P}} - \ton{\boldsymbol{\mathrm{P}}\boldsymbol\cdot\boldsymbol\nabla}\boldsymbol{\mathrm{Q}}$,  with $\boldsymbol{\mathrm{P}}\rightarrow {\boldsymbol{\mathrm{B}}}_\mathrm{g},\,\boldsymbol{\mathrm{Q}}\rightarrow\boldsymbol r$ so that $\boldsymbol\nabla\boldsymbol\cdot\boldsymbol r = 3,\,\ton{{\boldsymbol{\mathrm{B}}}_\mathrm{g}\boldsymbol\cdot\boldsymbol\nabla}\boldsymbol\cdot\boldsymbol r= {\boldsymbol{\mathrm{B}}}_\mathrm{g}$,
and the assumption that ${\boldsymbol{\mathrm{B}}}_\mathrm{g}$ is uniform yielding $\boldsymbol\nabla\boldsymbol\cdot{\boldsymbol{\mathrm{B}}}_\mathrm{g} = 0$ and $ \ton{\boldsymbol r\boldsymbol\cdot\boldsymbol\nabla}{\boldsymbol{\mathrm{B}}}_\mathrm{g} = 0$.}
\eqi
-4{\boldsymbol{\mathrm{A}}}_\mathrm{g} = {\boldsymbol{\mathrm{B}}}_\mathrm{g}\boldsymbol\times\boldsymbol r,\lb{Olar2}
\eqf
occurs as if it were affected by the Coriolis acceleration\footnote{Indeed, \rfr{ACor}, calculated with \rfr{OLar} and \rfr{Bg}, yields just the time--honored Lense--Thirring acceleration \citep{iers10}. In this case, the gravitomagnetic field of \rfr{Bg} is not uniform; the equivalence with the Coriolis acceleration felt in a rotating frame is, thus, just local.}
\eqi
{\boldsymbol{\mathcal{A}}}_\mathrm{C}= 2\boldsymbol{v}\boldsymbol{\times}{\boldsymbol{\Omega}}_\mathrm{L}\lb{ACor}
\eqf
experienced in a noninertial frame rigidly rotating with angular velocity
\eqi
{\boldsymbol{\Omega}}_\mathrm{L} := \rp{{\boldsymbol{\mathrm{B}}}_\mathrm{g}}{2c}.\lb{OLar}
\eqf
By means of \rfr{hacca}, \rfr{Olar2} and \rfr{OLar}, \rfr{Ace} can be cast just into the form\footnote{The penultimate step of \rfr{Larmo} is explained by the uniformity hypothesis of ${\boldsymbol{\mathrm{B}}}_\mathrm{g}$.}
\eqi
\boldsymbol A= c\derp{\boldsymbol{\mathfrak{h}}}{t} = \rp{2}{c}\derp{\boldsymbol{\mathrm{A}}_\mathrm{g}}{t} = -\rp{1}{2c}\derp{{\boldsymbol{\mathrm{B}}}_\mathrm{g}}{t}\boldsymbol\times\boldsymbol{r} = -\rp{1}{2c}\dert{{\boldsymbol{\mathrm{B}}}_\mathrm{g}}{t}\boldsymbol\times\boldsymbol{r} = -\dert{{\boldsymbol\Omega}_\mathrm{L}}{t}\boldsymbol\times\boldsymbol r\lb{Larmo}.
\eqf
Thus, the gravitational analogue of the Larmor's theorem holds exactly also when the gravitomagnetic field of the source, assumed spatially uniform, is explicitly time--dependent \citep{2002IJMPD..11..781I}.

Aim of the paper is the calculation of the impact of \rfr{Accel}, viewed as a small correction to the dominant Newtonian inverse--square term, on the orbital motion of a satellite of the spinning body by assuming a purely precessional\footnote{The case of a linearly time--dependent  $J\ton{t}$, with the primary's spin unit vector $\boldsymbol{\hat{J}}$ aligned with the reference $z$ axis,  was treated in \citet{2010GReGr..42.2393R}.} motion of $\boldsymbol J$; such a task is accomplished in full generality in Section~\ref{calcolo}. Numerical evaluations of the size of the resulting effects on the motion of the spacecraft Juno currently orbiting Jupiter and of the double pulsar PSR J0737--3039 A/B are given in Section~\ref{Junosec} where also the case of a star orbiting a supermassive Kerr black hole is treated. \textcolor{black}{Section~\ref{fine} summarizes the findings  and offers conclusions.}
\section{The averaged rates of change of the Keplerian orbital elements}\lb{calcolo}
The net rates of change of the Keplerian orbital elements of a test particle revolving about the spinning primary can be  analytically worked out by means of the Gauss equations, valid for any perturbing acceleration $\boldsymbol A$
\begin{align}
\dert{a}{t} \lb{dadt} &= \rp{2}{\nk \sqrt{1-e^2}} \qua{e A_{r} \sin f + \ton{\rp{p}{r}} A_{\tau}},\\ \nonumber \\
\dert e t \lb{dedt} & = \rp{\sqrt{1-e^2}}{\nk a} \grf{A_{r} \sin f + A_{\tau} \qua{\cos f + \rp{1}{e} \ton{1-\rp{r}{a}} }}, \\ \nonumber\\
\dert I t \lb{dIdt}& = \rp{1}{\nk a \sqrt{1-e^2}} A_{h} \ton{\rp{r}{a}} \cos u, \\ \nonumber\\
\dert \Omega t \lb{dOdt}& = \rp{1}{\nk a \sin I \sqrt{1-e^2}} A_{h} \ton{\rp{r}{a}} \sin u, \\ \nonumber\\
\dert \omega t \lb{dodt} & = \rp{\sqrt{1-e^2}}{\nk a e} \qua{-A_{r} \cos f + A_{\tau} \ton{1 + \rp{r}{p}} \sin f} - \cos I \dert\Omega t, \\ \nonumber \\
\dert\eta{t} \lb{detadt} &= -\rp{2}{\nk a} A_{r} \ton{\rp{r}{a}} - \rp{\ton{1-e^2}}{\nk a e} \qua{-A_{r} \cos f +A_{\tau} \ton{1+\rp{r}{p}} \sin f}.
\end{align}
In \rfrs{dadt}{detadt}, $a$ is the semimajor axis, $e$ is the eccentricity, $\nk:=\sqrt{GM/a^3}$ is the Keplerian mean motion, $p:= a\ton{1 - e^2}$ is the semilatus rectum, $I$ is the inclination of the orbital plane to the reference plane adopted, $\Omega$ is the longitude of the ascending node, $\omega$ is the argument of pericentre, $\eta$ is the mean anomaly at epoch, $f$ is the true anomaly, and $u:=\omega + f$ is the argument of latitude. Furthermore, $A_r,A_\tau,A_h$ are the projections of the perturbing acceleration at hand onto the radial, transverse and normal directions, respectively determined by the mutually orthonormal vectors
\begin{align}
\boldsymbol{\hat{r}} \lb{ur} & = \grf{\cos\Omega\cos u - \cos I\sin\Omega\sin u,\,\sin\Omega\cos u + \cos I\cos\Omega\sin u,\,\sin I\sin u},\acap
\boldsymbol{\hat{\tau}} \lb{ut} & = \grf{ -\cos\Omega\sin u - \cos I\sin\Omega\cos u,\,-\sin\Omega\sin u + \cos I\cos\Omega\cos u,\,\sin I\,\cos u},\acap
\boldsymbol{\hat{h}} \lb{un} & = \grf{\sin I\sin\Omega,\,-\sin I\cos\Omega,\,\cos I};
\end{align}
$\boldsymbol{\hat{h}}$ is normal to the orbital plane, being directed along the orbital angular momentum.
\Rfr{ur} can be also expressed as
\eqi
\boldsymbol{\hat{r}} = \boldsymbol{\hat{l}}\cos u + \boldsymbol{\hat{m}}\sin u.\lb{er}
\eqf
In \rfr{er},
\begin{align}
\boldsymbol{\hat{l}} \lb{elle}& := \grf{\cos\Omega,~\sin\Omega,~0},\acap
\boldsymbol{\hat{m}} \lb{emme}& := \grf{-\cos I\sin\Omega,~\cos I\cos\Omega,~\sin I}
\end{align}
are two unit vectors lying in the orbital plane; $\boldsymbol{\hat{l}}$ is directed along the line of nodes, while $\boldsymbol{\hat{m}}$ is perpendicular to $\boldsymbol{\hat{l}}$ in such a way that
\eqi
\boldsymbol{\hat{l}}\boldsymbol{\times}\boldsymbol{\hat{m}} = \boldsymbol{\hat{h}}.
\eqf
The right--hand--sides of \rfrs{dadt}{detadt}, calculated for the disturbing acceleration under consideration,  are to be first evaluated onto the Keplerian ellipse
\eqi
r = \rp{p}{1 + e\cos f},
\eqf
assumed as unperturbed reference trajectory, and then integrated over one full orbital revolution of the test particle by means of
\eqi
\dert t f = \rp{r^2}{\sqrt{GM p}};
\eqf
the averaged rates are finally obtained by dividing the resulting expressions for the aforementioned integrations by the Keplerian orbital period $\Pb = 2\uppi/\nk$.

By assuming a purely precessional motion for  the spin angular momentum of the primary, i.e. for
\eqi
\dert{\boldsymbol J}t = {\boldsymbol{\Omega}}_\mathrm{p}\boldsymbol\times\boldsymbol J,\lb{rotaz}
\eqf
where ${\boldsymbol{\Omega}}_\mathrm{p}$ is the precession velocity vector of $\boldsymbol J$,
and by means of the Binet--Cauchy identity \citep[Eq. (25), pag. 76]{Gibbs01}
\eqi
\ton{\boldsymbol{\mathrm{C}}\boldsymbol\times\boldsymbol{\mathrm{D}}}\boldsymbol\cdot\ton{\boldsymbol{\mathrm{E}}\boldsymbol\times\boldsymbol{\mathrm{F}}} = \ton{\boldsymbol{\mathrm{C}}\boldsymbol\cdot\boldsymbol{\mathrm{E}}}\ton{\boldsymbol{\mathrm{D}}\boldsymbol\cdot\boldsymbol{\mathrm{F}}} - \ton{\boldsymbol{\mathrm{C}}\boldsymbol\cdot\boldsymbol{\mathrm{F}}}\ton{\boldsymbol{\mathrm{D}}\boldsymbol\cdot\boldsymbol{\mathrm{E}}},\lb{bin}
\eqf
the radial, transverse and normal components of \rfr{Accel}
can be finally cast into the form
\begin{align}
A_r \lb{AR} & = 0, \acap
A_\tau & = \rp{2GJK_1}{c^2r^2},\acap
A_h \lb{AN} & = -\rp{2GJ\ton{K_2\cos u + K_3\sin u}}{c^2r^2},
\end{align}
where
\begin{align}
K_1 \lb{K1}&:= \ton{{\boldsymbol{\Omega}}_\mathrm{p}\boldsymbol\times\boldsymbol{\hat{J}}}\boldsymbol\cdot\boldsymbol{\hat{h}}, \acap
K_2 \lb{K2}& := \ton{{\boldsymbol{\Omega}}_\mathrm{p}\boldsymbol\cdot\boldsymbol{\hat{h}}}\ton{\boldsymbol{\hat{J}}\boldsymbol\cdot\boldsymbol{\hat{l}}} - \ton{{\boldsymbol{\Omega}}_\mathrm{p}\boldsymbol\cdot\boldsymbol{\hat{l}}}\ton{\boldsymbol{\hat{J}}\boldsymbol\cdot\boldsymbol{\hat{h}}}, \acap
K_3 \lb{K3}& := \ton{{\boldsymbol{\Omega}}_\mathrm{p}\boldsymbol\cdot\boldsymbol{\hat{h}}}\ton{\boldsymbol{\hat{J}}\boldsymbol\cdot\boldsymbol{\hat{m}}} - \ton{{\boldsymbol{\Omega}}_\mathrm{p}\boldsymbol\cdot\boldsymbol{\hat{m}}}\ton{\boldsymbol{\hat{J}}\boldsymbol\cdot\boldsymbol{\hat{h}}}.
\end{align}

By assuming that $\boldsymbol{\hat{J}}$ stays approximately constant during an orbital revolution, the integration of the right--hand--sides of \rfrs{dadt}{detadt}, calculated with \rfrs{AR}{AN}, straightforwardly yields
\begin{align}
\dert a t \lb{adot} & = \rp{4G J K_1}{c^2 \nk a^2\ton{1 - e^2}}, \acap
\dert e t \lb{edot} & = \rp{2G J\ton{1 - \sqrt{1 - e^2}}K_1}{c^2\nk a^3 e}, \acap
\dert I t \lb{Idot} & = \rp{G J\grf{K_2\qua{-e^2 + \ton{-2 + e^2 + 2 \sqrt{1 - e^2}}\cos 2\omega} + K_3\ton{-2 + e^2 + 2 \sqrt{1 - e^2}}\sin 2\omega}}{c^2\nk a^3 e^2\sqrt{1 - e^2}}, \acap
\dert \Omega t \lb{Odot} & = -\rp{G J\csc I\grf{K_3\qua{e^2 + \ton{-2 + e^2 + 2 \sqrt{1 - e^2}}\cos 2\omega} - K_2\ton{-2 + e^2 + 2 \sqrt{1 - e^2}}\sin 2\omega}}{c^2\nk a^3 e^2\sqrt{1 - e^2}}, \acap
\dert \omega t \lb{odot} & = \rp{G J\cot I\grf{K_3\qua{e^2 + \ton{-2 + e^2 + 2 \sqrt{1 - e^2}}\cos 2\omega} - K_2\ton{-2 + e^2 + 2 \sqrt{1 - e^2}}\sin 2\omega}}{c^2\nk a^3 e^2\sqrt{1 - e^2}}, \acap
\dert \eta t \lb{etadot}& = 0.
\end{align}
According to \rfrs{adot}{etadot},  all the orbital elements experience generally nonvanishing secular variations, apart from the mean anomaly at epoch whose precession is identically zero. Furthermore, \rfrs{adot}{etadot} have a general validity since they hold for any orbital configuration of the satellite, and for an arbitrary orientation of the primary's spin axis as well.
From \rfr{K1} and \rfrs{dadt}{dedt} it turns out that the rates of change of the semimajor axis and the eccentricity vanish for equatorial orbits, i.e. if $\boldsymbol{\hat{J}} = \pm\boldsymbol{\hat{h}}$. On the contrary, the precessions of the inclination, the node and the pericentre do not vanish in such a scenario, as per \rfr{rotaz}, \rfrs{K2}{K3} and \rfrs{Idot}{odot}. If the orbit is polar, i.e., if $\boldsymbol{\hat{J}}\boldsymbol\cdot\boldsymbol{\hat{h}} = 0 $, all the rates of the orbital elements do not vanish provided that ${\boldsymbol{\Omega}}_\mathrm{p}$ does not lie in the orbital plane; in this case, the inclination, the node and the pericentre stay constant, as per \rfrs{K2}{K3} and \rfrs{Odot}{odot}.

In the limit of small eccentricities, \rfrs{adot}{etadot} reduce to
\begin{align}
\dert a t \lb{adot0} & = \rp{4G J K_1}{c^2 \nk a^2} + \mathcal{O}\ton{e^2}, \acap
\dert e t \lb{edot0} & = \rp{e G J K_1}{c^2\nk a^3} + \mathcal{O}\ton{e^2}, \acap
\dert I t \lb{Idot0}& = -\rp{G J K_2}{c^2\nk a^3} + \mathcal{O}\ton{e^2}, \acap
\dert \Omega t \lb{Odot0} & = -\rp{G J\csc I K_3}{c^2\nk a^3} + \mathcal{O}\ton{e^2}, \acap
\dert \omega t \lb{odot0} & = \rp{G J\cot I K_3}{c^2\nk a^3}  + \mathcal{O}\ton{e^2}. \acap
\dert \eta t \lb{etadot0}& = 0.
\end{align}
It turns out that, to the lowest order in $e$, the semimajor axis, the inclination, the node and the pericentre formally undergo secular variations even for circular orbits, while the rate of the eccentricity is proportional to $e$ itself.

In a previous work \citep{2002IJMPD..11..781I}, only the rates of change of the semimajor axis, the eccentricity, the inclination and the node were calculated, to the zeroth order in $e$, in the particular case of the LAGEOS satellite \citep{1985JGR....90.9217C} orbiting the Earth whose spin axis undergoes the lunisolar precession.
\section{Numerical evaluations for the Juno--Jupiter and the double pulsar PSR J0737--3039 A/B systems}\lb{Junosec}
\subsection{Juno and Jupiter}
According to \citet{2016P&SS..126...78L}, the Jupiter's pole is precessing due to the gravitational tugs of the Sun, its satellites and the other bodies of the solar system about the normal ${\boldsymbol{\hat{w}}}_0$ to the Sun--Jupiter invariable plane which can be approximately assumed equal to the solar system's invariable plane \citep{2012A&A...543A.133S}.

The unit vector ${\boldsymbol{\hat{w}}}_0$ can be expressed as \citep{2012A&A...543A.133S}
\eqi
{\boldsymbol{\hat{w}}}_0 = \grf{\sin i_\mathrm{p}\sin\theta_\mathrm{p},\,-\sin i_\mathrm{p}\cos\theta_\mathrm{p},\,\cos i_\mathrm{p}},\lb{w0}
\eqf
where \citep{2012A&A...543A.133S}
\begin{align}
i_\mathrm{p} &\simeq 23^\circ, \acap
\theta_\mathrm{p} &\simeq 3.8^\circ;
\end{align}
such figures are referred to the International Celestial Reference Frame (ICRF) having the mean Earth's equator at epoch as reference plane.

The Jovian spin axis can be parameterized as
\begin{align}
\boldsymbol{\hat{J}} = \grf{\cos\alpha\cos\delta,\,\sin\alpha\cos\delta,\,\sin\delta},
\end{align}
where the nominal values of the R.A. $\alpha$ and decl. $\delta$ of the Jupiter's pole are approximately \citep{2016P&SS..126...78L}
\begin{align}
\alpha & \simeq 268^\circ,\acap
\delta & \simeq 64^\circ.
\end{align}

The spin precession rate of Jupiter is about \citep{2016P&SS..126...78L}
\eqi
\Omega_\mathrm{p}\simeq 3700\,\mathrm{mas\,yr}^{-1},\lb{Omp}
\eqf
where $\mathrm{mas\,yr}^{-1}$ stands for milliarcseconds per year.

By calculating \rfr{un} with the values
\begin{align}
I & = 92.99^\circ,\acap
\Omega & = 267.52^\circ
\end{align}
for the inclination and the node of the Juno spacecraft \citep{2017SSRv..213....5B}, referred to the ICRF,  retrieved from the web interface HORIZONS, maintained by the NASA Jet Propulsion Laboratory (JPL), and inserting \rfrs{w0}{Omp} in \rfr{K1}, one obtains
\eqi
K_1\simeq -2.5\times 10^{-14}\,\mathrm{s}^{-1} = -0.00005^\circ\,\mathrm{yr}^{-1}.\lb{K1J}
\eqf

Thus, the rate of the semimajor axis of Juno, obtained from \rfr{adot} calculated with \rfr{K1J}, the value of the Jovian angular momentum
\eqi
J \simeq  6.9\times 10^{38}\,\mathrm{kg\,m}^2\,\mathrm{s}^{-1}
\eqf
reported in \citet{2003AJ....126.2687S}, and the figures
\begin{align}
a & = 4.06\times 10^6\,\mathrm{km}, \acap
e & = 0.981
\end{align}
for the semimajor axis and the eccentricity of the spacecraft retrieved from HORIZONS, turns out to be
\eqi
\dert a t \simeq -2\,\mathrm{\mu m\,yr}^{-1}.
\eqf

The rate of change of the eccentricity can be calculated  with \rfr{edot} in the same way as just done for the semimajor axis getting
\eqi
\dert e t \simeq -1.5\,\mathrm{pas\,yr}^{-1},\lb{edotJ}
\eqf
where pas yr$^{-1}$ stands for picoarcseconds per year.

Since the calculated values of $K_2$ and $K_3$ are similar to that of \rfr{K1J}, the precessions of the inclination, the node and the perijove of Juno are of the same order of magnitude of \rfr{edotJ}.
\subsection{The double pulsar}
The spin angular momentum ${\boldsymbol{J}}_\mathrm{B}$ of the member B of the double pulsar PSR J0737--3039 A/B \citep{2003Natur.426..531B,2004Sci...303.1153L} undergoes a precession\footnote{It is the general relativistic geodetic or de Sitter precession of the  spin of an object moving in the deformed spacetime of a nonspinning massive body \citep{DamRu74,1975PhRvD..12..329B}.} of $4.77^\circ\,\mathrm{yr}^{-1}$ \citep{2008Sci...321..104B}. The orbital period of the binary is as short as $\Pb = 2.45\,\mathrm{hr}$ \citep{2006Sci...314...97K}, while the spin period of B amounts to $T_\mathrm{B} = 2.77\,\mathrm{s}$ \citep{2006Sci...314...97K}; thus, by assuming for its moment of inertia the standard value
\eqi
\mathcal{I}\simeq 1\times 10^{38}\,\mathrm{kg\,m}^2
\eqf
for neutron stars \citep{LorKra05}, the size of its angular momentum is of the order of
\eqi
J_\mathrm{B} \simeq 2.2\times 10^{38}\,\mathrm{kg\,m}^2\,\mathrm{s}^{-1}.
\eqf

It turns out that the semimajor axis rate amounts to, at most,
\eqi
\left|\dert a t\right|\lesssim 0.1\,\mathrm{mm\,yr}^{-1},
\eqf
while the orbital precessions are at the $\simeq 0.1-10\,\mathrm{nas\,yr}^{-1}$ (nanoarcseconds per year) level.
\subsection{A supermassive black hole--star scenario}
The effects under examination might become relevant in the case of, say, a star orbiting a spinning supermassive black hole (SMBH) whose spin axis $\boldsymbol{\hat{J}}_\bullet$ undergoes, for some reasons\footnote{It may happen, e.g., in supermassive black hole binaries \citep{2008PhRvD..78d4021R,2021MNRAS.501.2531S}.}, a relatively fast precession.

Let a hypothetical test particle orbits a SMBH  at a distance of, say, 100 Schwarzschild radii\footnote{The Schwarzschild radius of a black hole of mass $M_\bullet$ is $\mathcal{R}_\mathrm{S} := 2GM_\bullet/c^2$.}
along a\textcolor{black}{n almost} circular orbit. \textcolor{black}{It should be recalled} that the magnitude of the angular momentum a Kerr black hole \citep{1963PhRvL..11..237K,2015CQGra..32l4006T} is \citep{1986bhwd.book.....S}
\eqi
J_\bullet = \chi\rp{M_\bullet^2 G}{c},
\eqf
with $\left|\chi\right|\leq 1$\textcolor{black}{. If} one assumes, say,
\begin{align}
M_\bullet &= 4.5\times 10^6 M_\odot, \acap
\chi & = 1,
\end{align}
where $M_\odot$ is the Sun's mass, it turns out that\textcolor{black}{, according to \rfr{adot0},} the particle's semimajor axis
would be changed by at most \eqi
\rp{1}{a}\left|\dert a t\right|\lesssim 7\%\,\mathrm{yr}^{-1},
\eqf
provided that the hole's spin axis precessional frequency $\Omega_\mathrm{p}$ is $10\%$ of the star's orbital one $\nk$, i.e., if
\eqi
\rp{\Omega_\mathrm{p}}{\nk}=0.1.\lb{Opnk}
\eqf
\textcolor{black}{By relying upon the same assumptions, \rfrs{Idot0}{odot0} tell that the precessions of the other orbital elements would be}
\eqi
\textcolor{black}{\left|\dert \kappa t\right|\lesssim 1^\circ\,\mathrm{yr}^{-1},\,\kappa =I,\Omega, \omega}
\eqf
\textcolor{black}{while the rate of change of the eccentricity would be of the order of}
\eqi
\textcolor{black}{\left|\dert e t\right|\lesssim 10^{-6}\,\mathrm{yr}^{-1},}
\eqf
\textcolor{black}{as per \rfr{edot0}. Since \rfrs{adot0}{etadot0} are proportional to $\Omega_\mathrm{p}$, the previous values can be straightforwardly rescaled for different values of \rfr{Opnk}. As far as the distance from the black hole is concerned, both the relative rate of change of $a$ and \rfrs{edot0}{odot0} fall as $a^{-3/2}$.  }

\section{Summary and conclusions}\lb{fine}
The impact of the general relativistic Euler--type gravitomagnetic acceleration induced by the temporal variation of mass--energy currents  was analytically calculated, to the first post--Newtonian order, for a restricted two--body system in the hypothesis that the source of the gravitational field is an isolated, massive body rigidly rotating whose spin angular momentum undergoes a purely precessional motion.

The calculation has a full generality since it holds for any orbital configuration of the test particle, and for an arbitrary orientation of the precession velocity vector of the central object as well.

It turns out that, in general, all the orbital elements, apart from the mean anomaly at epoch, undergo long--term variations calculated by assuming that the primary's spin axis precession is much slower than the satellite's orbital revolution.

The resulting effects are usually quite small; for the Juno spacecraft currently orbiting Jupiter, the semimajor axis changes at a rate as small as a few microns per year, while the shifts of  the other orbital elements are at the picoarcseconds per year level. For the double pulsar, the resulting figures are larger by just a few orders of magnitude. On the other hand, they may become relevant around a supermassive black hole. Indeed,  by assuming a precessional frequency of the hole's spin axis equal to $10\%$ of the mean motion of a putative test particle orbiting it at 100 Schwarzschild radii, it turns out that its semimajor axis changes by up to $7\%$ per year if the mass of the hole at the Galactic Centre is assumed.
\section*{Data availability}
No new data were generated or analysed in support of this research.
\section*{Conflict of interest statement}
I declare no conflicts of interest.
\bibliography{Megabib}{}
\end{document}